\documentclass{PoS}

\usepackage{epsf}
\title{The strange quark content of the nucleon in 2+1 flavor lattice QCD}

\ShortTitle{2+1 flavor nucleon strangeness}

\author{\speaker{Walter Freeman}\\
        University of Arizona\\
        E-mail: \email{wfreeman@physics.arizona.edu}}

\author{Doug Toussaint\\
        University of Arizona\\
        E-mail: \email{doug@physics.arizona.edu}}

\def\LL{\left\langle}	
\def\RR{\right\rangle}	
\def\PAR#1#2{ {{\partial #1}\over{\partial #2}} }

\newcommand{\BE}{\begin{displaymath}}
\newcommand{\EE}{\end{displaymath}}
\newcommand{\BNE}{\begin{equation}}
\newcommand{\ENE}{\end{equation}}
\newcommand{\BEA}{\begin{eqnarray}}
\newcommand{\EEA}{\nonumber\end{eqnarray}}

\abstract{The strangeness of the nucleon, $\LL N| \bar s s |N\RR - \LL 0| \bar s s |0\RR$ , is a quantity of
interest for interpreting
the results of dark matter detection experiments as well as for exploring
the structure of the nucleon itself. We present a calculation of this quantity
in 2+1 flavor lattice QCD using a range of lattice spacings and quark masses. The method
is based on
calculating quark-line disconnected contributions on the MILC lattice
configurations, which include the effects of dynamical strange quarks.
After continuum and chiral extrapolations, the value is $\LL N| \bar s s |N\RR - \LL 0| \bar s s |0\RR =
0.69 \pm 0.07_{stat} \pm 0.09_{sys}$ in the $\overline{ms} (2 GeV)$
regularization.}

\FullConference{The XXVII International Symposium on Lattice Field Theory - LAT2009\\
		 July 26-31 2009\\
		 Peking University, Beijing, China}

\begin{document}
\section {Introduction and motivations}

The strange-quark content of the nucleon is difficult to measure experimentally but is a quantity of wide interest. 
In particular, the interaction cross section between some proposed dark matter candidates (for instance, neutralinos)
and ordinary matter may 
have a large contribution from interactions with sea strange quarks in the nucleon.
Specifically, the interest is in the quantity $\LL N| \int\, d^3x\, \bar s s |N\RR - \LL 0| \int\, d^3x\, \bar s s |0\RR$:
the connected part of the strange quark condensate, integrated over the volume of the nucleon.
Knowledge of this matrix element is crucial to design experimental schemes for dark matter detection and to interpret
their results. If the matrix element is known, it is possible to determine the constraints on the parameter space of 
dark matter candidates accessible to a given experiment. \cite{BALTZ06,ELLIS08}

As the quantity in question cannot be measured experimentally or calculated perturbatively, 
we must turn to lattice QCD to compute it.
Previous lattice calculations to answer this question have been done with quenched simulations\cite{FUKUGITA02,DONG02},
with a 2-flavor sea using Wilson or overlap quarks\cite{SESAM98,UKQCD01,JLQCD08}, and 2+1 flavor stout quarks\cite{BALI08}.
Recently this quantity was extracted from baryon mass fits to 2+1 flavor simulations \cite{YOUNGTHOMAS09}. The
results from these studies often have large uncertainties, and some conclude that $\LL N | \bar s s | N \RR$ may be
significantly larger than its natural size of unity.

We outline a method for calculating this quantity by evaluating disconnected quark-line diagrams,
and present the results of applying this method to the MILC Collaboration's library of
gauge configurations using 2+1 flavors of Asqtad-improved staggered quarks. An additional very long
ensemble of 4566 lattices from the UKQCD Collaboration with $\beta = 6.75$, $m_l = 0.06a$, $m_s = 0.30a$, $a \sim 0.125$ fm
using the same action is included, allowing a
measurement of $\LL N | \bar s s | N \RR$ using 25788 gauge configurations using a full 2+1 flavor sea~\cite{THISWORK}.

The method outlined here can be used for light quarks as well, and for hadrons other than the nucleon.
While measurements of their quark condensates are not as immediately in demand, they may provide
useful information about chiral perturbation theory low-energy constants,
and may lead to further progress in understanding the structure of the QCD sea within hadrons. 

\section {The MILC lattice generation program}

The MILC collaboration is engaged in an extensive project of QCD simulations
using a Symanzik-improved gauge action and the Asqtad-improved staggered-quark action with a 2+1 flavor sea.
This project consists of a number of runs at nominal lattice spacings 
of $a=.12$~fm, $a=.09$~fm, and $a=.06$~fm, 
along with other coarser runs not considered here. 
Details of the action, the ensembles of gauge configurations, and the 
method of extracting nucleon correlators can be found in Ref.~\cite{RMP}. An additional very long
ensemble of 4566 lattices from UKQCD with $\beta = 6.75$, $m_l = 0.06a$, $m_s = 0.30a$ 
using the same action is included in the analysis.

\section {The method}

In a chiral fermion formulation, the matrix element in question is equal to
$\PAR{M_N}{m_s}$ by the Feynman-Hellman theorem. We emphasize that the derivative 
should be taken holding all other parameters in the action fixed. A sizeable value 
for ${\partial M_N}\over{\partial m_q}$ does not imply 
a large dependence of the {\it physical} nucleon mass on $m_s$.
Changing the value of $m_s$ changes the value of {\it all} physical quantities by a similar amount, and this change
is interpreted as an overall rescaling of the lattice rather than a shift in the physical values.
Likewise, a large result does not imply the presence of many $\bar s s$ loops in the nucleon.
Rather, it comes from the suppression of the vacuum $\bar s s$ condensate
near the nucleon.

One needs large-scale simulations with a true 2+1 flavor sea to determine 
$\LL N|s \overline{s} |N \RR$ with reasonable accuracy. 
Previous methods used to determine this quantity require particular
choices for the lattice parameters, namely ensembles with different $m_s$ but the same $\beta$.
On the other hand, in a general program of lattice QCD simulations such as MILC's, 
$m_s$ is typically held fixed at its physical value while $\beta$ is changed with the $m_q$'s to keep the physical lattice spacing fixed. 
Thus, computing $\LL N|s \overline{s} |N \RR$ using the standard method requires special-purpose simulations 
which must be limited in scope due to economics. We have developed a method to determine 
${\partial M_N}\over{\partial m_s}$ from any {\it single} lattice ensemble, allowing its use
on the MILC lattice configurations.
The many ensembles are simply used to conduct an extrapolation to the physical point and to improve statistics.

The nucleon mass $M_N$ is obtained by a fit to the nucleon correlator $C(t)$ and as such can be thought of as a complicated
function of the correlator at different times: $M_N = f(C(t_1),C(t_2),C(t_3)...)$. The crucial idea is that one can use the 
chain rule for differentiation to rewrite the derivative:

\BNE
{{\partial M_N}\over{\partial m_q}} = {{\partial M_N} \over {\partial C(t_1)}} {{\partial C(t_1)} \over {\partial m_s}}
+ {{\partial M_N}\over{\partial C(t_2)}} {{\partial C(t_2)} \over {\partial m_s}}
+ {{\partial M_N}\over{\partial C(t_3)}} {{\partial C(t_3)} \over {\partial m_s}...} \ENE

The partial derivatives ${\partial M_N}\over{\partial C(t_i)}$ can be evaluated most simply by applying a small perturbation
to the nucleon correlator and examining the change in the fit result. The other partial derivative ${\partial P(t_i)}\over{\partial m_s}$ can be evaluated by an application of the Feynman-Hellman
theorem in reverse to relate it to $\LL P(t_i) \overline {s} s \RR - \LL P(t_i) \RR \LL \overline {s} s \RR$.
%
%

Here we take advantage of the fact that, whenever the MILC code generates or reads a lattice for analysis,
it prints a stochastic estimator for $\int \, d^4x \, \bar s s$. The number of estimators used in this work, 6-16 
per lattice,
is sufficient for their fluctuation to contribute significantly to the statistical error. These
values can be used to compute $\LL C(t) \int \, d^4x \, \bar s s \RR - \LL C(t) \RR \LL \int \, d^4x \, \bar s s \RR$
by simple evaluation with no additional use of computer time. By doing this for
each $t$ used in the fit to determine $M_N$, it is possible to evaluate the chain rule sum and determine 
${\partial M_N}\over{\partial m_s}$.

This double use of the Feynman-Hellman theorem to access the nuclear strangeness might seem redundant: 
as we have the values of $\LL \overline {s} s \RR$,
why not simply evaluate the matrix element $\LL N|s \overline{s} |N \RR$ directly? The problem is that the lattice operator
used to create and annihilate the nucleon overlaps with many other three-quark states; 
the nucleon is simply the lowest-lying three-quark state. 
Furthermore, the normalization of the nucleon state created this way is unknown. 
Using this operator along with the $\LL \overline {s} s \RR$ data to evaluate $\LL N|s \overline{s} |N \RR$ 
directly would give the sea quark content of some superposition of states of unknown normalization, {\it not}
of the nucleon itself. The fitting procedure used to extract the nucleon mass provides a way of
extracting information about the nucleon alone, and the double use of the Feynman-Hellman theorem
provides a way to relate it to $\LL N|s \overline{s} |N \RR$.

Statistical errors on the result for ${\partial M_N}\over{\partial m_s}$ were calculated using
the jackknife method with blocks of size 10; this is large enough to take into account autocorrelations
in the lattice data. Use of larger jackknife blocks produces an insignificant change in the size of the
errors. 

\section {Choice of fit range}

Choosing a lower minimum distance in the nucleon mass fits will result in lower statistical error but
may introduce systematic biases from pollution by excited states. The minimum fit distance required to avoid significant 
systematic error in $\LL N|s \overline{s} |N \RR$ can be smaller than the one used for precision measurements of $M_N$. 
When extracting $M_N$, any pollution of the correlator by the excited states 
in the fit range will cause an incorrectly high value for the mass. However, the lowest-lying
excited state is the delta. As we expect the effect of the delta on the strange quark condensate 
to be broadly similar to that of the nucleon, a small amount of delta pollution will not create substantial systematic error.
 
\begin{figure*}[tbh]
\hspace{-0.1in}
\includegraphics[width=0.325\textwidth]{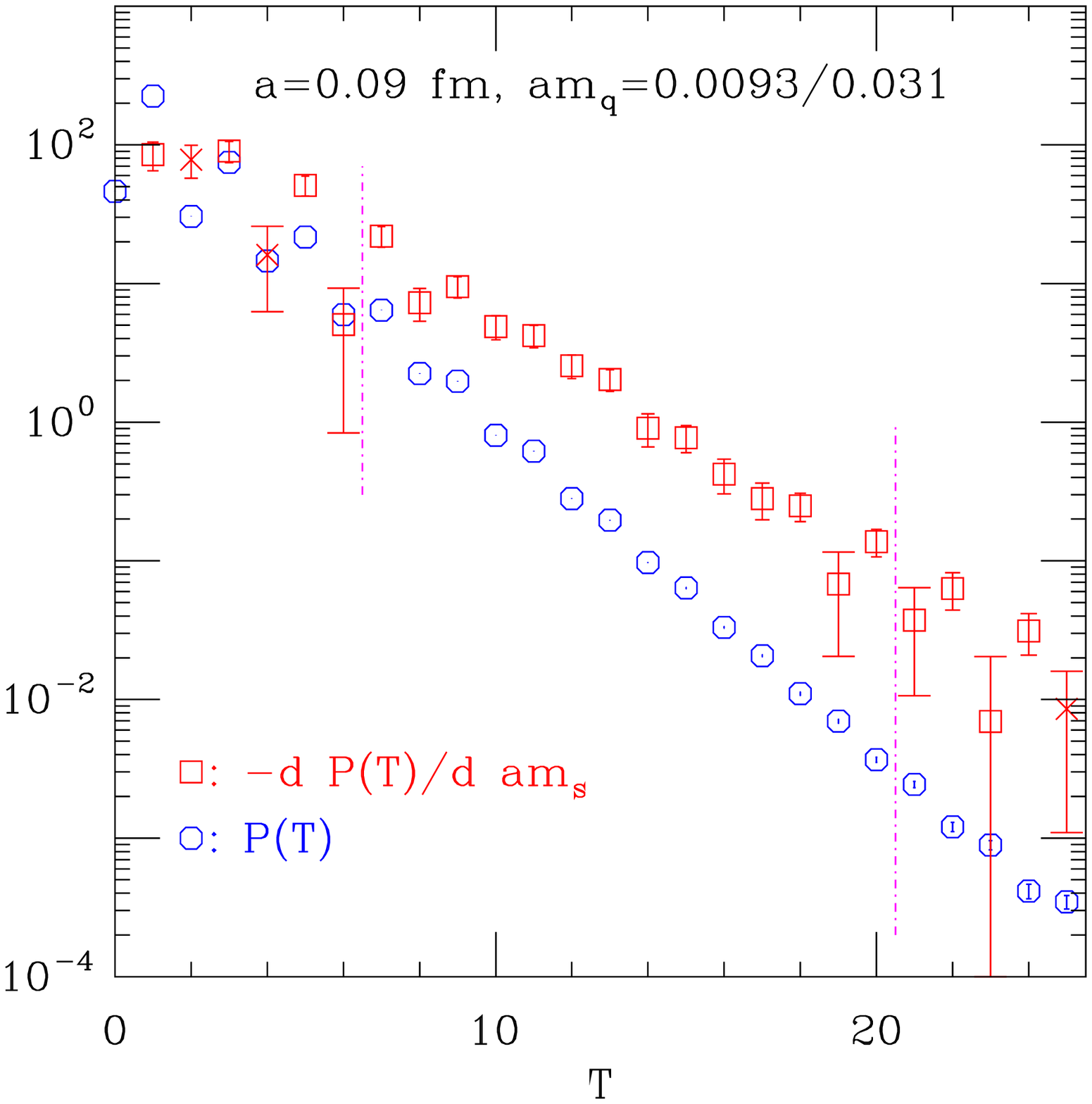}
\includegraphics[width=0.335\textwidth]{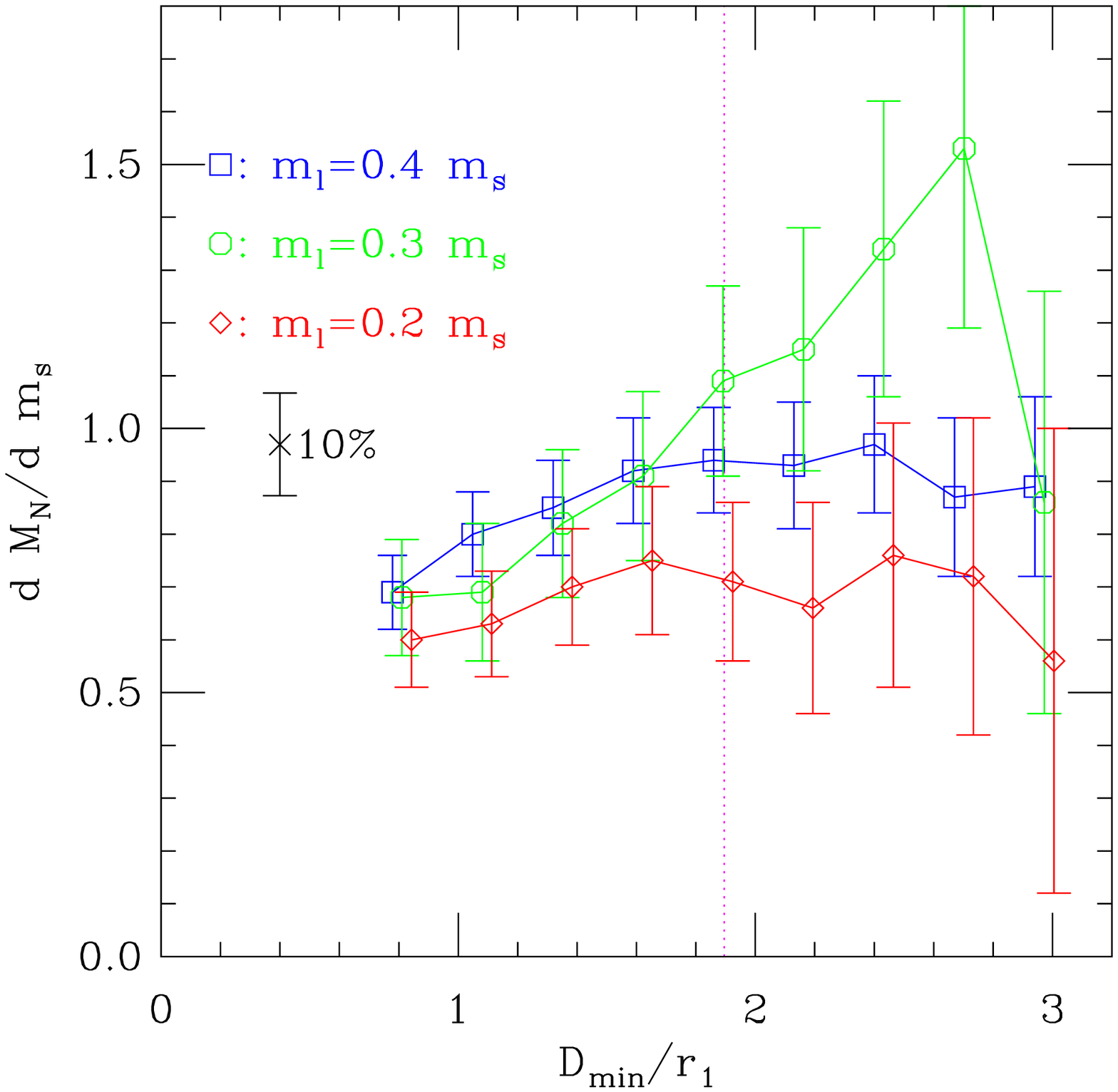}
\includegraphics[width=0.325\textwidth]{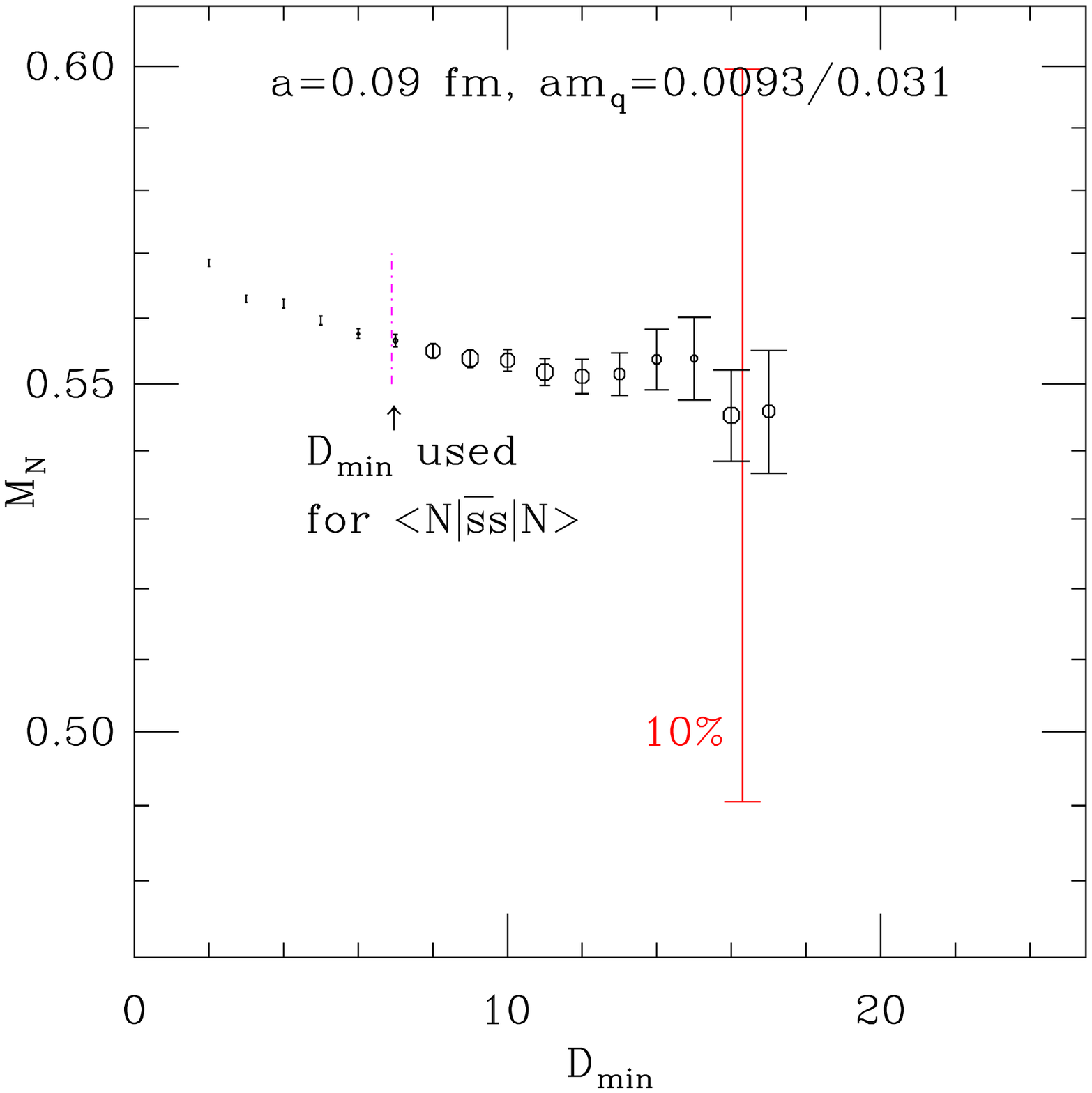}
\caption{The nucleon correlator and the derivative
of this correlator with respect to $m_s$ for the ensemble with
$am_l=0.0093$ and $am_s=0.031$ (first panel).  For the derivative, the squares
are points where the derivative is negative, and crosses are points
where it is positive.  The vertical lines show the range used in fitting the correlator.
The second panel shows $\PAR{M_N}{m_s}$ for three ensembles with $a\approx 0.9$fm as a
function of the minimum distance used in the fitting, and the third panel shows
the fitted nucleon mass itself versus $t_{min}$.  The error bars labelled ``10\%''
in the second and third panels show the size of the ten percent
systematic error estimate from excited state contamination.
\label{samplefigs}
}
\end{figure*}

The minimum distances chosen should be consistent 
in physical units between lattice spacings. The minimum distance chosen should also
be in a region where the result at successive minimum distances does not differ greatly. Both looking at
the result ensemble-by-ensemble and looking at fits to all ensembles in the same nominal lattice spacing suggest the use of
$t_{min} = 0.6$~fm. This choice is also suggested by the signal-to-noise ratio in $\PAR{C(t)}{m_s}$; see Figure~\ref{samplefigs}. In lattice units, this gives 
$t_{min} = 5a$ ($a=0.13$fm), $t_{min} = 7a$ ($a=0.09$fm), and $t_{min} = 10a$ ($a=0.06$fm). 
By examining the dependence of $\PAR{M_N}{m_s}$ on $t_{min}$, we conservatively estimate the systematic
error due to excited state pollution as 10\%. However, the nucleon mass itself can be computed
with much lower statistical error at higher $t_{min}$; mass fits at these higher minimum distances differ from those at $t_{min} = 0.6$~fm 
by only 1\% to 5\%. Thus 10\% is potentially an overly-pessimistic estimate.

The result is quite insensitive to the maximum distance
used for the fits; this is expected, since the signal-to-noise ratio of the correlator is very poor there.

\begin{figure}[tbh]
\hspace{-0.1in}
\begin{center}
\includegraphics[width=.4\textwidth]{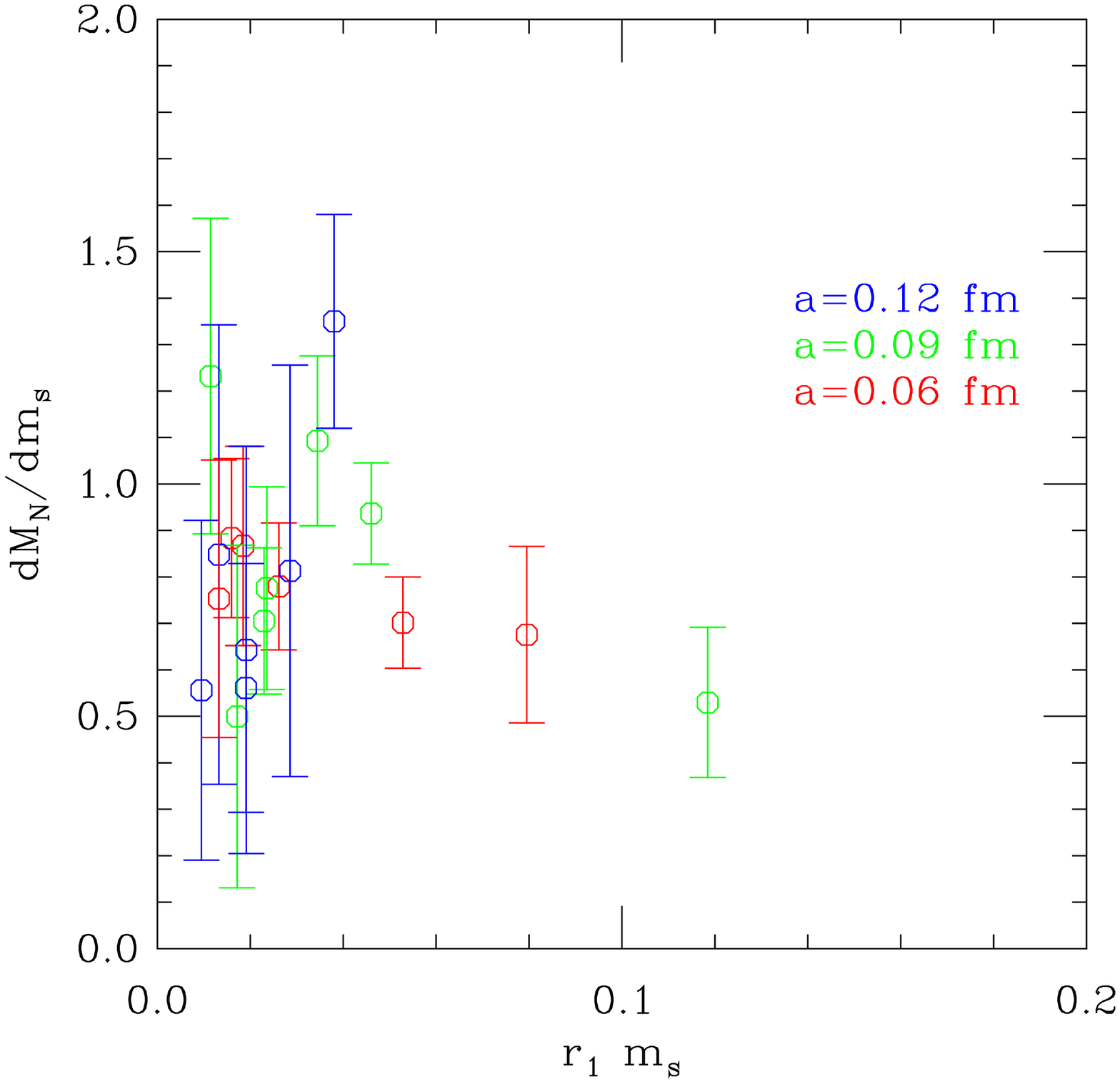}
\includegraphics[width=.4\textwidth]{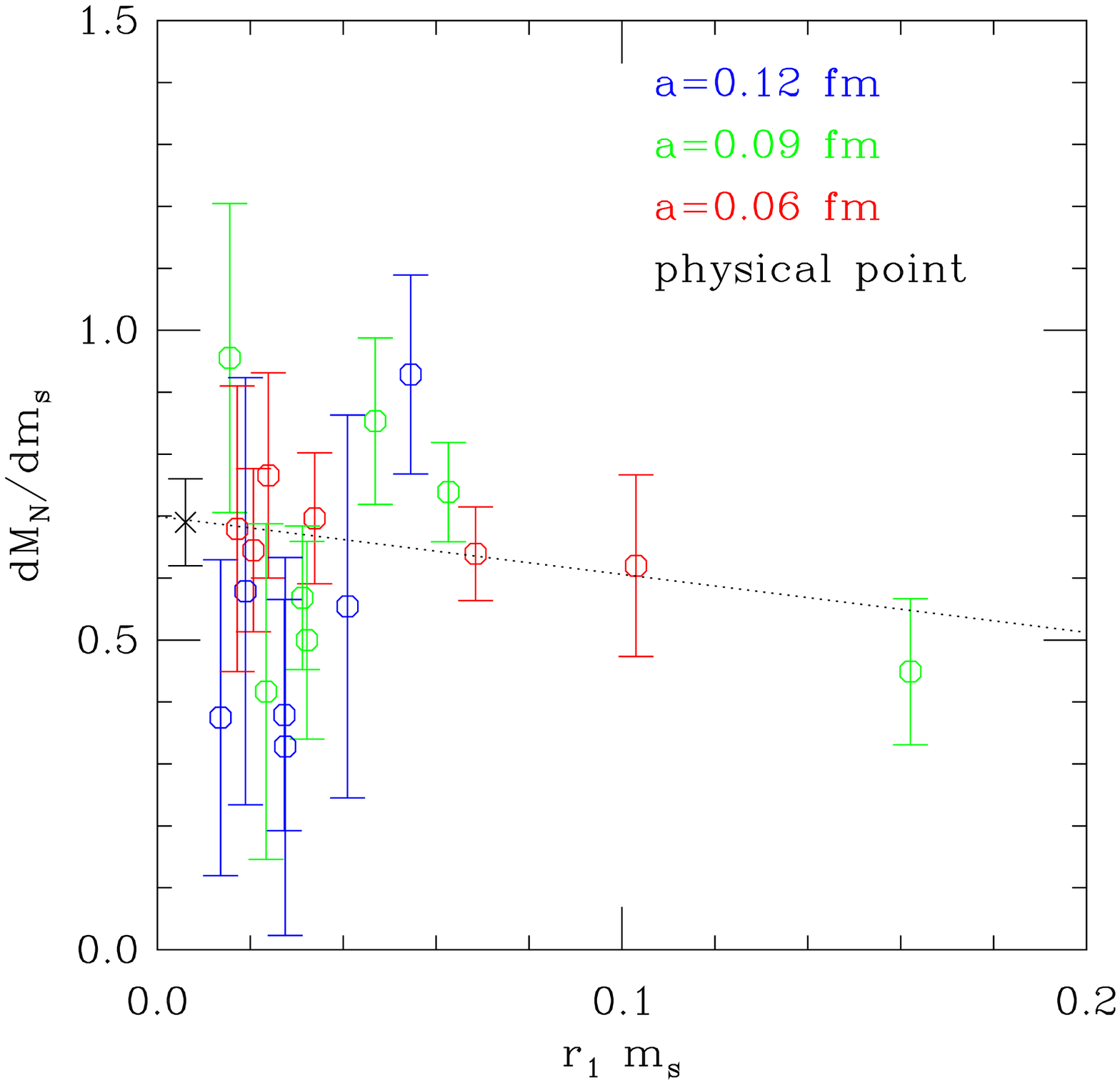}
\end{center}
\caption{$\PAR{M_N}{m_s}$ evaluated on the MILC ensembles.
The first panel shows the ``bare'' results; the second panel shows the
results converted to the $\overline{ms}({\rm 2\ GeV})$ renormalization and 
adjusted to the physical strange quark mass. The dotted line shows the chiral
and continuum fit; the black point at $m_l = .005$ shows this fit evaluated at the physical point.}
\label{resultfigs}
\end{figure}

The values of ${\partial M_N}\over{\partial m_s}$ obtained on each ensemble using these minimum distances are tabulated here and shown
in the first panel of Figure~\ref{resultfigs}. In the table, both the bare value (in the lattice regularization), and the value
converted to the $\overline{ms}({\rm 2\ GeV})$ regularization and shifted to the correct strange quark mass (see below) are given.

\begin{center}
\small
\renewcommand{\arraystretch}{0.35}
\begin{tabular}{|c | c c | c | c | c | c |}
\hline
 $\beta$ & $a m_l$ & $a m_s$ & $a$ (fm) &$N_{lats}$ & $\LL N \bar s s N \RR$ (bare) & $\LL M \bar s s N \RR$ (adj) \\
\hline
  6.81&   0.30 &  0.50& 0.117 &    552  &  0.676(190) & 0.620(147)\\
  6.79&   0.20 &  0.50& 0.118 &     2067&  0.702(98) & 0.639(76)\\
  6.76&   0.10 &  0.50& 0.119 &  2278&  0.779(137) & 0.696(106)\\
  6.76&   0.07 &  0.50& 0.118 & 2098&  0.867(214) & 0.766(166)\\
  6.76&   0.05 &  0.50& 0.117 & 2033&  0.753(299) & 0.679(230)\\
  6.75&   0.06 &  0.30& 0.117 & 4566&  0.884(171) & 0.645(132)\\
\hline
  7.08& 0.0031 & 0.031& 0.084 & 1013 &  1.232(339) & 0.955(249)\\
  7.085& 0.00465& 0.031& 0.084 &    599  &  0.500(369) & 0.417(271)\\
  7.09& 0.0062&  0.031& 0.084 & 1943&  0.705(158) & 0.568(116)\\
  7.10& 0.0093&  0.031& 0.084 &   1137  &  1.093(183) & 0.853(134)\\
  7.11& 0.0124&  0.031& 0.084 & 1993&  0.936(109) & 0.739(80)\\
  7.18& 0.0310&  0.031& 0.081 &    496  &  0.530(161) & 0.449(118)\\
  7.10& 0.0062&  0.0186&0.082 &    948  &  0.776(218) & 0.500(159)\\
\hline
  7.46& 0.0018&  0.018 & 0.059 &   823  &  0.556(366) & 0.375(255)\\
 7.465& 0.0025&  0.018  & 0.059 &  798   &  0.848(494)& 0.579(345) \\
 7.47 & 0.0036&  0.018 & 0.058 &658     & 0.561(268) & 0.379(187)\\
 7.475& 0.0054&  0.018 & 0.059 &616     & 0.813(443) & 0.554(309)\\
 7.48 & 0.0072&  0.018 & 0.059 &620     & 1.352(249) & 0.929(160)\\
 7.46 & 0.0036&  0.0108& 0.058 &550     & 0.643(438) & 0.328(305)\\
\hline
\end{tabular}
\end{center}

\section{Analysis}

\subsection{Renormalization}

Since in the continuum extrapolation it is necessary to compare quantities measured at 
different lattice spacings and thus different regularization schemes, it is necessary to
convert all of the data to a common renormalization scheme since $\PAR{m_N}{m_s}$ is a 
renormalization dependent quantity. Since the final result will be presented in the $\overline{MS}$
renormalization scheme at a scale of 2 GeV so it will be most useful to the broader
physics community, it is preferable to begin the analysis by converting all values to 
this renormalization scale. The Z-factors for this conversion are known ~\cite{QUARKMASS2}. 

\subsection{Strange quark mass dependence}

Since the value of $m_s$ in lattice units is not known prior to a lattice run, 
all of the ensembles were run at strange quark masses different than the 
physical one; on some ensembles this error is large (20\%). It is thus necessary to determine the 
dependence of $\PAR{m_N}{m_s}$ on $m_s$ (that is, $\frac {\partial^2 m_N}{\partial m_s^2}$)
to perform an extrapolation to the physical $m_s$.

As MILC has run one or two ensembles at each nominal lattice spacing
with $m_s$ set to 60\% of the guess at the physical value,
it should be possible to determine $\frac {\partial^2 m_N}{\partial m_s^2}$ by examining 
$\PAR{m_N}{m_s}$ on both these ensembles and those with the heavier strange quarks. However,
since these ``light strange'' ensembles are short, better results
for this second derivative can be obtained by a different method.

On many ensembles, the light quark is quite heavy; in some cases, it is 
actually closer to the physical value of $m_s$ than the lattice heavy quark. Thus the behavior of the ``light'' quark condensate on these ensembles is
similar to that of a strange quark condensate with a lower strange quark mass.

These ensembles with a heavier light quark allow for the measurement of the 
behavior of a ``heavy quark condensate'' at two different quark masses, and allow us to
estimate the second derivative 
$\frac {\partial^2 M_N}{\partial m_s^2} = (\frac {\partial M_N}{\partial m_s}-\frac {\partial M_N}{\partial m_l})/(m_s-m_l)$.
Since $\PAR{M_N}{m_s}$ and $\PAR{M_N}{m_l}$ are measured on the same lattices, 
their values are correlated, so the error on $\PAR{M_N}{m_s}-\PAR{M_N}{m_l}$ (determined by jackknife) is reduced.

By examining five ensembles with a large number of configurations and relatively heavy light
quarks, it is possible to determine $\frac {\partial^2 M_N}{\partial m_s^2} = -2.2$ by a weighted
average of its value on each, estimated as above. This value is then used to extrapolate each
data point to $m_{s,phys}$. The second panel of Figure~\ref{resultfigs} shows the data converted to $\overline{MS} (2 GeV)$ and adjusted to the correct $m_s$.

\subsection{Light quark mass dependence and continuum extrapolation}

Similarly, the value of $\PAR{M_N}{m_s}$ may depend on $m_l$, and we are most interested
in evaluating it at the physical light quark mass. In this case
the data set contains results for $\PAR{M_N}{m_s}$ at many different values of $m_l$, so
it is possible to determine the dependence on $m_l$ with a simple fit. Examination of the
$\chi$PT form for $\PAR{M_N}{m_s}$ reveals a constant plus linear fit in $m_l$ is good enough;
no $\chi$PT terms at higher-order are relevant at the level of statistical
accuracy provided by the present data~\cite{FRINKMEISSNER}.

It is also necessary to extrapolate to the continuum. In
the Asqtad fermion formulation, the leading-order errors in the action are proportional to $a^2$.
Thus, the leading-order effect on $\PAR{M_N}{m_s}$ will likewise
be proportional to $a^2$; this effect can be determined by adding such a term to the fit form. 
However, such a term will be poorly constrained, since the better statistics in our dataset are
from the $a=.12$ fm ensembles, and since the effect from lattice spacing is small. Other hadronic
quantities calculated using the MILC Asqtad data show roughly a 10\% effect between the coarse lattices
and the continuum, so we use a Bayesian prior with a central value of 0 and a width 
corresponding to a 10\% effect to constrain the lattice spacing dependent term in the fit.

As both the light quark mass dependence and the lattice spacing dependence will be computed in the
same fit, the proper fit form is $\PAR{M_N}{m_s} = A + B m_l + C a^2$, with a Bayesian constraint on C.

\section{Result and error budget}

Evaluating the fit above at the physical value of $m_l$ and in the continuum, we find $\PAR{M_N}{m_s} = 0.69 \pm 0.07_{statistical}$.
We estimate the systematic error due to excited states present in the nucleon correlator,
as discussed above, at 10\%. The extrapolation to the physical light quark mass involves higher order terms in $\chi$PT which were not considered here. To estimate the size of this effect, we note that if the nucleon mass itself is fit to 
a constant-plus-linear form over the range considered here, the result is seven percent off from the result obtained when two more
orders in $m_{\pi}$ are added to the fit. We thus estimate the effect of higher order terms in ${\chi}$PT as 7\%. In one case where
a spatial volume larger than the one used here has been run, the nucleon mass computed on the larger volume differs by 1\%. Since
the effect on the strange quark condensate is potentially more sensitive to finite volume effects, we estimate the systematic error
due to finite volume effects as 3\%. Finally, Ref.~\cite{QUARKMASS2} quotes an error in $Z_m$ as 4\%. If these errors are combined in
quadrature we thus estimate the total systematic error as 0.09.

The renormalization-invariant quantity $m_s \PAR{M_N}{m_s}$ is also commonly quoted. Using a similar fit, we calculate $m_s \PAR{M_N}{m_s} = 59(6)(8)$ MeV.
This quantity does not include uncertainty in $Z_m$, as this cancels, but includes a lattice systematics error of nearly the same size, coming from
the 2\% uncertainty in the lattice scale and the uncertainty in the lattice strange quark mass.

\end{document}